\documentclass[11pt]{article}

\usepackage{times}
\usepackage[margin=1in]{geometry}
\usepackage{amsmath,amssymb,amsthm}
\usepackage{latexsym}
\usepackage{rotating}
\usepackage{url}
\usepackage{hyperref}
\usepackage{amsrefs}

\newcommand{\captionfonts}{\small}
\makeatletter
\long\def\@makecaption#1#2{%
  \vskip\abovecaptionskip
  \sbox\@tempboxa{{\captionfonts #1: #2}}%
  \ifdim \wd\@tempboxa >\hsize
    {\captionfonts #1.~~#2\par}
  \else
    \hbox to\hsize{\hfil\box\@tempboxa\hfil}%
  \fi
  \vskip\belowcaptionskip}
\makeatother

\newcommand{\ab}{\allowbreak}

\DeclareMathOperator{\Tr}{Tr}
\DeclareMathOperator{\ADV}{ADV}
\DeclareMathOperator{\MADV}{ADV^\pm}
\DeclareMathOperator{\Toeplitz}{Toeplitz}
\newcommand{\OSP}{\mathrm{OSP}}
\newcommand{\advmax}{\substack{\Gamma \ge 0 \\ \Gamma \ne 0}}
\newcommand{\floor}[1]{{\left\lfloor #1 \right\rfloor}}
\newcommand{\ceil}[1]{{\left\lceil #1 \right\rceil}}
\newcommand{\R}{\mathbb{R}}
\newcommand{\Z}{\mathbb{Z}}
\newcommand{\g}{\gamma}

\def\01{\{0,1\}}
\newcommand{\norm}[1]{\| #1 \|}
\newcommand{\sn}[1]{\| #1 \|}
\newcommand{\trn}[1]{\| #1 \|_{\Tr}}

\newtheorem{theorem}{Theorem}
\newtheorem{corollary}[theorem]{Corollary}
\newtheorem{lemma}[theorem]{Lemma}
\newtheorem{proposition}[theorem]{Proposition}

\newcommand{\eqnref}[1]{(\ref{#1})}
\newcommand{\figref}[1]{Figure~\ref{#1}}
\newcommand{\secref}[1]{Section~\ref{#1}}
\newcommand{\thmref}[1]{Theorem~\ref{#1}}
\newcommand{\corref}[1]{Corollary~\ref{#1}}
\newcommand{\lemref}[1]{Lemma~\ref{#1}}
\newcommand{\propref}[1]{Proposition~\ref{#1}}

\newcommand{\be}{\begin{equation}}
\newcommand{\ee}{\end{equation}}
\def\ba#1\ea{\begin{align}#1\end{align}}
\def\bas#1\eas{\begin{align*}#1\end{align*}}

\newcommand{\lab}[1]{\mbox{\scriptsize$#1$}}
\newcommand{\sw}[1]{\begin{sideways}\lab{#1}\end{sideways}}
\newlength{\newcolsep}
\newlength{\oldwidth}
\newcommand{\labelmatrix}[3]{%
\settowidth{\newcolsep}{#2}
\setlength{\newcolsep}{.5\newcolsep}
\settowidth{\oldwidth}{\sw{00001111}}
\addtolength{\newcolsep}{-.5\oldwidth}
\addtolength{\newcolsep}{\arraycolsep}
\begin{array}{r@{}c@{}l}
& \setlength{\arraycolsep}{\newcolsep} \begin{matrix}
      \sw{11110000} & \sw{01111000} & \sw{00111100} & \sw{00011110} 
    & \sw{00001111} & \sw{10000111} & \sw{11000011} & \sw{11100001}
 \end{matrix} & \\
#1 & \begin{bmatrix}#3\end{bmatrix} &
\begin{matrix}
\lab{11110000}\\
\lab{01111000}\\
\lab{00111100}\\
\lab{00011110}\\
\lab{00001111}\\
\lab{10000111}\\
\lab{11000011}\\
\lab{11100001}
\end{matrix}
\end{array}}

\makeatletter
\def\revddots{\mathinner{\mkern1mu\raise\p@ 
\vbox{\kern7\p@\hbox{.}}\mkern2mu 
\raise4\p@\hbox{.}\mkern2mu\raise7\p@\hbox{.}\mkern1mu}} 
\makeatother

\begin{document}

\title{Optimal quantum adversary lower bounds for ordered search}
\author{Andrew M.\ Childs\thanks{Department of Combinatorics \& Optimization and Institute for Quantum Computing, University of Waterloo; amchilds@uwaterloo.ca}
\and
Troy Lee\thanks{Laboratoire de Recherche en Informatique, Universit{\'e} Paris-Sud; troyjlee@gmail.com}}
\date{}
\maketitle


\begin{abstract}
The goal of the ordered search problem is to find a particular item in an ordered list of $n$ items.  Using the adversary method, H{\o}yer, Neerbek, and Shi proved a quantum lower bound for this problem of $\frac{1}{\pi} \ln n + \Theta(1)$.  Here, we find the exact value of the best possible quantum adversary lower bound for a symmetrized version of ordered search (whose query complexity differs from that of the original problem by at most $1$).  Thus we show that the best lower bound for ordered search that can be proved by the adversary method is $\frac{1}{\pi} \ln n + O(1)$.  Furthermore, we show that this remains true for the generalized adversary method allowing negative weights.
\end{abstract}

\section{Introduction}

Search is a fundamental computational task.  In a general search problem, one is looking for a distinguished item in a set, which may or may not have some structure.  At one extreme, in the \emph{unstructured search problem}, we assume the set has no additional structure whatsoever.  In this setting, a classical search requires $\Omega(n)$ queries in the worst case to find the distinguished item.  Grover's well-known search algorithm shows that a quantum computer can find the distinguished item with high probability in only $O(\sqrt{n})$ queries \cite{Gro97}.  A lower bound based on a precursor to the adversary method shows this is optimal up to a constant factor \cite{BBBV97}.

At the other extreme of search problems, in the \emph{ordered search problem}, we assume our set comes equipped with a total order, and we are able to make comparison queries, i.e., queries of the form `$w \le z$?'.  Classically, we can apply binary search to find the desired item in $\ceil{\log_2 n}$ queries, and an information theoretic argument shows this is tight.

Quantum computers can speed up the solution of the ordered search problem by a constant multiplicative factor.  Farhi, Goldstone, Gutmann, and Sipser developed a class of translation-invariant ordered search algorithms and showed that one such algorithm, applied recursively, gives an exact ordered search algorithm using $3 \log_{52} n \approx 0.526 \log_2 n$ quantum queries \cite{FGGS99}.
Brookes, Jacokes, and Landahl used a gradient descent search to find an improved translation-invariant algorithm, giving an upper bound of $4\log_{550} n \approx 0.439 \log_2 N$ queries \cite{BJL04}.
Childs, Landahl, and Parrilo used numerical semidefinite optimization to push this approach still further, improving the upper bound to $4\log_{605} n \approx 0.433 \log_2 n$ \cite{CLP07}.
Ben-Or and Hassidim gave an algorithm based on adaptive learning that performs ordered search with probability of error $o(1)$ using only about $0.32 \log_2 n$ queries \cite{BH07}.

In fact, the quantum speedup for ordered search is not more than a constant multiplicative factor.
Using the quantum adversary method \cite{Amb02}, H{\o}yer, Neerbek, and Shi showed a lower bound of $\frac{1}{\pi}(\ln n -1) \approx 0.221 \log_2 n$ queries \cite{HNS02}, improving on several previous results \cites{Amb99,BW99,FGGS98}.  However, the exact value of the best possible speedup factor, a fundamental piece of information about the power of quantum computers, remains undetermined.

In this paper, we give some evidence that the asymptotic quantum query complexity of ordered search is $\frac{1}{\pi}\ln n +O(1)$.  Specifically, we show that the best lower bound given by the adversary method, one of the most powerful techniques available for showing lower bounds on quantum query complexity, is $\frac{1}{\pi}\ln n +O(1)$.  We show this both for the standard adversary method \cite{Amb02} and the recent strengthening of this method to allow negative weights \cite{HLS07}.  In particular, we prove the following: 

\begin{theorem}
\label{thm:ospadv}
Let $\ADV(f)$ be the optimal bound given by the adversary method for the function $f$, let $\MADV(f)$ be the optimal value of the adversary bound with negative weights, and let $\OSP_n$ the ordered search problem on $n$ items (symmetrized as discussed in \secref{sec:advosp}).  Then
\bas
  \ADV(\OSP_{2m})
  &=2 \sum_{i=0}^{m-1} \left(\frac{\binom{2i}{i}}{4^i}\right)^2 \\
  \ADV(\OSP_{2m+1})
  &=2 \sum_{i=0}^{m-1} \left(\frac{\binom{2i}{i}}{4^i}\right)^2
    +\left(\frac{\binom{2m}{m}}{4^m}\right)^2.
\eas
Furthermore,
$$
  \MADV(\OSP_{n}) \le \ADV(\OSP_n) + O(1).
$$
\end{theorem}

The bounds described in \thmref{thm:ospadv} are asymptotically $\frac{2}{\pi} \ln n + O(1)$, but are always strictly larger than the H{\o}yer-Neerbek-Shi bound.  Understanding the best possible adversary bound for small $n$ could be useful, since the best exact algorithms for ordered serach have been found by discovering a good algorithm for small values of $n$ and using this algorithm recursively.  Furthermore, since the adversary quantity can be viewed as a simplification of the quantum query complexity, we hope that our analytic understanding of optimal adversary bounds will provide tools that are helpful for determining the quantum query complexity of ordered search.

The remainder of this article is organized as follows.  In \secref{sec:adversary}, we briefly review the quantum adversary method.  In \secref{sec:osp}, we define the basic ordered search problem as well as a extended version that is more symmetric, and hence easier to analyze.  In \secref{sec:advosp}, we apply the adversary method to the symmetrized ordered search problem and present semidefinite programs characterizing it, both in primal and dual formulations.  In \secref{sec:nonneg}, we find the optimal non-negative adversary lower bound for ordered search and compare it to the bound of \cite{HNS02}.  Then we show in \secref{sec:neg} that negative weights do not substantially improve the bound.  Finally, we conclude in \secref{sec:conclusion} with a brief discussion of the results.

\section{Adversary bound}
\label{sec:adversary}

The adversary method, along with the polynomial method \cite{BBCMW01}, is one of the two main techniques for proving lower bounds on quantum query complexity. The adversary method was originally developed by Ambainis \cite{Amb02}, with roots in the hybrid method of \cite{BBBV97}.  It has proven to be a versatile technique, with formulations given by various authors in terms of spectral norms of matrices \cite{BSS03}, weight schemes \cites{Amb06,Zha05}, and Kolmogorov complexity \cite{LM04}. \v{S}palek and Szegedy showed that all these versions of the adversary method are in fact equivalent \cite{SS06}.  Recently, H{\o}yer, Lee, and \v{S}palek developed a new version of the adversary method using negative weights which is always at least as powerful as the 
standard adversary method, and can sometimes give better lower bounds \cite{HLS07}.

We will use the spectral formulation of the adversary bound, as this version best expresses the similarity between the standard and negative adversary methods.  In this formulation, the value of the adversary method for a function $f$ is given by
$$
\ADV(f):=\max_{\advmax} \frac{\sn{\Gamma}}{\max_i \sn{\Gamma \circ D_i}},
$$
where $\Gamma$ is a square matrix with rows and columns indexed by the possible inputs $x \in S \subseteq \01^n$, constrained to satisfy $\Gamma[x,y]=0$ if $f(x)=f(y)$;  $D_i$ is a zero/one matrix with $D_i[x,y]=1$ if $x_i \ne y_i$ and $0$ otherwise; $A \circ B$ denotes the Hadamard (i.e., entrywise) product of matrices $A$ and $B$; and $\Gamma \ge 0$ means that the matrix $\Gamma$ is entrywise non-negative.  Note that the set $S$ of possible inputs need not be the entire set $\01^n$ of all $n$-bit strings---in other words, $f$ might be a partial function, as is the case for ordered search.

The negative adversary method is of the same form, but removes the restriction to non-negative matrices in the maximization.  Thus the value of the negative adversary method for a function $f$ is given by
$$
\MADV(f):=\max_{\Gamma \ne 0} \frac{\sn{\Gamma}}{\max_i \sn{\Gamma \circ D_i}}.
$$

The relation of these adversary bounds to quantum query complexity is given by
the following theorem.
Let $Q_\epsilon(f)$ denote the minimum number of quantum queries to $f$ needed to compute that function with error at most $\epsilon$.
Then we have

\begin{theorem}[\cites{Amb02,HLS07}]
Let $S\subseteq \01^n$, and let $\Sigma$ be a finite set.  Then for any function $f:S \rightarrow \Sigma$,
$$
  Q_{\epsilon}(f) \ge \frac{1-2\sqrt{\epsilon(1-\epsilon)}}{2}\ADV(f)
  \quad\text{and}\quad
  Q_{\epsilon}(f) \ge \frac{1-2\sqrt{\epsilon(1-\epsilon)}-2\epsilon}{2}\MADV(f).
$$
In particular, $Q_0(f) \ge \frac{1}{2}\MADV(f) \ge \frac{1}{2}\ADV(f)$.
\end{theorem}

\section{Ordered search problem}
\label{sec:osp}

In the ordered search problem, we are looking for a marked element $w$ in a set $Z$ equipped with a total order.  Let the members of $Z$ be $z_1 \le z_2 \le \cdots \le z_n$.  We are looking for the marked element $w \in Z$, and are able to ask queries of the form `$w \le z$?' for $z \in Z$.  Notice that if $w$ is the $i$th element in the list, then the answer to this query will be `no' for $z=z_j$ with $j<i$, and will be `yes' otherwise.  Thus we can model this problem as finding the first occurence of a `$1$' in a string $x \in \01^n$ where $x_j=0$ for $j<i$ and $x_j=1$ otherwise.  For example, for $n=4$, the possible inputs for the ordered search problem are $1111$, $0111$, $0011$, and $0001$, corresponding to the marked item being first, second, third, or fourth in the ordered set, respectively.  Thus we have transformed the input into a binary string, and the queries are to the bits of this input.  The goal is to determine which input we have---in other words, the function takes a different value on each input.

In general, when trying to determine the query complexity of a function $f$, it is helpful to consider its symmetries, as expressed by its \emph{automorphism group}.  We say that $\pi \in S_n$, a permutation of the $n$ bits of the input, is an automorphism of the function $f$ provided it maps inputs to inputs, and $f(x) = f(y) \Leftrightarrow f(\pi(x)) = f(\pi(y))$.  The set of automorphisms of any function on $n$-bit inputs is a subgroup of $S_n$, called the automorphsim group of that function.

The ordered search problem as formulated above has a trivial automorphism group, because any nontrivial permutation maps some input to a non-input.  However, we can obtain a more symmetric function, with only a small change to the query complexity, by putting the input on a circle \cite{FGGS99}.  Now let the inputs have $2n$ bits,  and consist of those strings obtained by cyclically permuting the string of $n$ $1$'s followed by $n$ $0$'s.  For example, with $n=4$, the inputs are $11110000,\ab 01111000,\ab 00111100,\ab 00011110,\ab 00001111,\ab 10000111,\ab 11000011,\ab 11100001$.  Again, we try to identify the input, so the function $\OSP_n$ takes a different value on each of the $2n$ inputs.  The automorphism group of $\OSP_n$ is isomorphic to $\Z_{2n}$, a fact that we will exploit in our analysis.

The query compexity of this extended function is closely related to that of the original function.  Given an $n$-bit input $x$, we can simulate a $2n$-bit input by simply querying $x$ for the first $n$ bits, and the complement of $x$ for the second $n$ bits.  In the other direction, to simulate an $n$-bit input using a $2n$-bit input, first query the $n$th bit of the $2n$-bit input.  If it is $1$, then we use the first half of the $2n$-bit input; otherwise we use the second half (or, equivalently, the complement of the first half).  Thus the query complexity of the extended function is at least that of the original function, and at most one more than that of the original function, a difference that is asymptotically negligible.

\section{Adversary bounds for ordered search}
\label{sec:advosp}

Finding the value of the adversary method is as an optimization problem.  To analyze the adversary bound for ordered search, we will use symmetry to simplify this problem.  The same simplification applies to both the standard and negative adversary bounds, so we treat the two cases simultaneously.

Suppose we are trying to design a good adversary matrix $\Gamma$, and are deciding what weight to assign the $(x,y)$ entry.  Intuitively, it seems that if $(x,y)$ and $(x',y')$ are related by an automorphism, then they should look the same to an adversary, and hence should be given the same weight.  The \emph{automorphism principle} states that there is an optimal adversary matrix with this property.  Although this principle does not provide any advice about what weight to give a particular pair $(x,y)$, it can vastly reduce the optimization space by indicating that the adversary matrix should possess certain symmetries.

\begin{theorem}[Automorphism principle \cite{HLS07}]
Let $G$ be the automorphism group of $f$.  Then there is an optimal adversary matrix $\Gamma$ satisfying $\Gamma[x,y]=\Gamma[\pi(x),\pi(y)]$ for all $\pi \in G$ and all pairs of inputs $x,y$.  Furthermore, if $G$ acts transitively on the inputs (i.e., if for every $x,y$ there is an automorphism taking $x$ to $y$), then the uniform vector (i.e., the vector with each component equal to $1$) is a principal eigenvector of $\Gamma$.
\label{thm:auto}
\end{theorem}

The automorphism group for the ordered search problem on a list of size $n$, extended to a circle of size $2n$ as discussed in the previous section, is isomorphic to $\Z_{2n}$, generated by the element $(1\,2\,3\ldots 2n)$ that cyclically permutes the list.  This group acts transitively on the inputs, so by the automorphism principle, the uniform vector is a principal eigenvector of the adversary matrix.  In addition, any pairs $(x,y)$ and $(x',y')$ that have the same Hamming distance are related by an automorphism.  Thus we may assume that the adversary matrix has at most $n$ distinct entries, and that the $(x,y)$ entry depends only on the Hamming distance between $x$ and $y$.  As all strings have the same Hamming weight, the Hamming distance between any pair is even.  We let $\Gamma[x,y]=\gamma_i$ when $x,y$ have Hamming distance $2i$.  For example, with $n=4$, we have
$$
\labelmatrix{\Gamma=\,}{$\g_2$}{%
0    & \g_1 & \g_2 & \g_3 & \g_4 & \g_3 & \g_2 & \g_1 \\
\g_1 & 0    & \g_1 & \g_2 & \g_3 & \g_4 & \g_3 & \g_2 \\
\g_2 & \g_1 & 0    & \g_1 & \g_2 & \g_3 & \g_4 & \g_3 \\
\g_3 & \g_2 & \g_1 & 0    & \g_1 & \g_2 & \g_3 & \g_4 \\
\g_4 & \g_3 & \g_2 & \g_1 & 0    & \g_1 & \g_2 & \g_3 \\
\g_3 & \g_4 & \g_3 & \g_2 & \g_1 & 0    & \g_1 & \g_2 \\
\g_2 & \g_3 & \g_4 & \g_3 & \g_2 & \g_1 & 0    & \g_1 \\
\g_1 & \g_2 & \g_3 & \g_4 & \g_3 & \g_2 & \g_1 & 0}
$$
Since all rows have the same sum, the uniform vector is indeed an eigenvector, corresponding to the eigenvalue $\g_n + 2\sum_{i=1}^{n-1} \g_i$.

Transitivity of the automorphism group also implies that all matrices $\Gamma \circ D_i$ have the same norm, so it is sufficient to consider $\Gamma \circ D_{2n}$.  Again considering the example of $n=4$, we have
$$
\Gamma \circ D_{2n} = \begin{bmatrix}
0    & 0    & 0    & 0    & \g_4 & \g_3 & \g_2 & \g_1 \\
0    & 0    & 0    & 0    & \g_3 & \g_4 & \g_3 & \g_2 \\
0    & 0    & 0    & 0    & \g_2 & \g_3 & \g_4 & \g_3 \\
0    & 0    & 0    & 0    & \g_1 & \g_2 & \g_3 & \g_4 \\
\g_4 & \g_3 & \g_2 & \g_1 & 0    & 0    & 0    & 0   \\
\g_3 & \g_4 & \g_3 & \g_2 & 0    & 0    & 0    & 0   \\
\g_2 & \g_3 & \g_4 & \g_3 & 0    & 0    & 0    & 0   \\
\g_1 & \g_2 & \g_3 & \g_4 & 0    & 0    & 0    & 0
\end{bmatrix}.
$$
This matrix consists of two disjoint, identical blocks, so its spectral norm is simply the spectral norm of one of those blocks.  In general, $\Gamma \circ D_{2n}$ consists of two disjoint blocks, where each block is an $n \times n$ symmetric Toeplitz matrix with first row equal to $(\g_n, \g_{n-1}, \ldots, \g_1)$, denoted $\Toeplitz(\g_n,\g_{n-1},\ldots,\g_1)$.  Thus via the automorphism principle we have reduced the adversary bound to the semidefinite program
\be
  \max ~ \g_n + 2\sum_{i=1}^{n-1} \g_i
  \quad\text{subject to}\quad
  \sn{\Toeplitz(\g_n, \g_{n-1}, \ldots, \g_1)}\le 1, ~ \gamma_i \ge 0
\tag{P}
\label{eq:primal}
\ee
in the case of non-negative weights, and 
\be
  \max ~ \g_n + 2\sum_{i=1}^{n-1} \g_i
  \quad\text{subject to}\quad
  \sn{\Toeplitz(\g_n, \g_{n-1}, \ldots, \g_1)}\le 1
\tag{P$^\pm$}
\label{eq:neg_primal}
\ee
in the case of the negative adversary method.
We emphasize that the automorphism principle ensures there is no loss of generality in considering adversary matrices of this form---this program has the same optimal value as the best possible adversary bound.

We will also use the duals of these semidefinite programs to show upper bounds on the values of the adversary methods.  Straightforward dualization shows that the dual of \eqref{eq:primal} is
\be
  \min \Tr(P)
  \quad\text{subject to}\quad
  P \succeq 0, ~
  \Tr_i(P) \ge 1 \text{~for~} i=0,\ldots,n-1,
\tag{D}
\label{eq:dual}
\ee
and that the dual of \eqref{eq:neg_primal} is
\be
  \min \Tr(P+Q)
  \quad\text{subject to}\quad
  P,Q \succeq 0, ~
  \Tr_i(P-Q) = 1 \text{~for~} i=0,\ldots,n-1
\tag{D$^\pm$}
\label{eq:neg_dual}
\ee
where $P \succeq 0$ means that the matrix $P$ is positive semidefinite.

In general, by a \emph{solution} of a semidefinite program, we mean a choice of the variables that satisfies the constraints, but that does not necessarily extremize the objective function.  If a solution achieves the optimal value of the objective function, we refer to it as an \emph{optimal solution}.

\section{Non-negative adversary}
\label{sec:nonneg}

In this section, we consider the standard adversary bound.  We first present the lower bound of H{\o}yer, Neerbek, Shi as applied to the symmetrized version of ordered search.  Then we construct an improved adversary matrix, giving a solution of \eqref{eq:primal} that achieves the bound stated in \thmref{thm:ospadv}.  Finally, we exhibit a solution to \eqref{eq:dual} with the same value, showing that our construction is optimal.

\subsection{H{\o}yer, Neerbek, Shi construction}

Within the framework described above, the lower bound of \cite{HNS02} can be given very simply.  Set $\g_i=0$ if $i > \floor{n/2}$ and $\g_i=1/(\pi i)$ otherwise.  This gives an objective function of
$$
  \frac{2}{\pi} \sum_{i=1}^{\floor{n/2}} \frac{1}{i}
  \sim
  \frac{2}{\pi} \ln n.
$$
Continuing our example with $n=4$, consider the matrix $\Gamma \circ D_{2n}$ under this choice of weights:
$$
\Gamma \circ D_{2n} =
\frac{1}{\pi} \begin{bmatrix}
0   & 0   & 0   & 0   & 0   & 0   & 1/2 & 1   \\
0   & 0   & 0   & 0   & 0   & 0   & 0   & 1/2 \\
0   & 0   & 0   & 0   & 1/2 & 0   & 0   & 0   \\
0   & 0   & 0   & 0   & 1   & 1/2 & 0   & 0   \\
0   & 0   & 1/2 & 1   & 0   & 0   & 0   & 0   \\
0   & 0   & 0   & 1/2 & 0   & 0   & 0   & 0   \\
1/2 & 0   & 0   & 0   & 0   & 0   & 0   & 0   \\
1   & 1/2 & 0   & 0   & 0   & 0   & 0   & 0
\end{bmatrix}.
$$
This matrix consists of four disjoint blocks, so its spectral norm is equal to the largest spectral norm of these blocks. In general, we have four disjoint nonzero blocks (and in the case of $n$ odd, two additional $2 \times 2$ zero blocks).  Each nonzero block is equivalent up to permutation to $1/\pi$ times $Z_\floor{n/2}$, where $Z_m$ is the \emph{half Hilbert matrix} of size $m \times m$, namely the Hankel matrix
$$
Z_m:=
\begin{bmatrix}
1           & \frac{1}{2} & \frac{1}{3} & \cdots      & \frac{1}{m} \\
\frac{1}{2} & \frac{1}{3} & \cdots      & \frac{1}{m} & 0           \\
\frac{1}{3} & \vdots      & \revddots   & 0           & 0           \\
\vdots      & \frac{1}{m} & \revddots   & \revddots   & \vdots      \\
\frac{1}{m} & 0           & 0           & \cdots      & 0
\end{bmatrix}.
$$
This may be compared with the usual Hilbert matrix, whose $(i,j)$ entry is $1/(i+j-1)$.  The spectral norm of any finite Hilbert matrix is at most $\pi$,  so as the half Hilbert matrix is non-negative and entrywise less than the Hilbert matrix, its spectral norm is also at most $\pi$.  (See the delightful article of Choi for this and other interesting facts about the Hilbert matrix \cite{Cho83}.)
This shows that the spectral norm of each matrix $\Gamma \circ D_i$ is at most $1$, giving a bound on the zero-error quantum query complexity of ordered search of approximately $\frac{1}{\pi}\ln n$.

\subsection{Optimal non-negative construction}

It turns out that one can do slightly better than the Hilbert weight scheme described above.  Here we construct the optimal solution to the adversary bound for $\OSP_n$ with non-negative weights.

A key role in our construction will be played by the sequence $\{\xi_i\}$, where
\be
  \xi_i:=\frac{\binom{2i}{i}}{4^i}.
\label{eq:xi}
\ee
This sequence has many interesting properties.
First, it is monotonically decreasing.  Consider the ratio
$$
\frac{\xi_{i+1}}{\xi_i}=\frac{\binom{2(i+1)}{i+1}4^i}{\binom{2i}{i} 4^{i+1}}
= \frac{2(i+1)(2i+1)}{4(i+1)^2}=\frac{i+1/2}{i+1} < 1.
$$
Indeed, this shows that $\{\xi_i\}$ is a hypergeometric sequence with the generating function
$$
  g(z) := \sum_{i=0}^\infty \xi_i z^i = {}_1F_0(\tfrac{1}{2};z) 
        = \frac{1}{\sqrt{1-z}}.
$$

These observations lead us to the next interesting property of our sequence.
\begin{proposition}
For any $j$,
$$
  \sum_{i=0}^j \xi_i \xi_{j-i} = 1.
$$
\label{prop:binom}
\end{proposition}

\begin{proof}
The product $g(z)^2$ is the generating function for the convolution appearing on the left hand side.  But $g(z)^2 = (1-z)^{-1}$, which has all coefficients equal to $1$, as claimed.
(For an alternative proof, using the fact that $\xi_i = (-1)^i \binom{-1/2}{i}$, see \cite{GKP89}*{p.~187}.)
\end{proof}

This proposition shows that the sequence $\{\xi_i\}$ behaves nicely under convolution.  We will also consider the behavior of $\{\xi_i\}$ under correlation.  Define
$$
  A_m(j):=\sum_{i=0}^{m-j-1}\xi_i \xi_{i+j}.
$$
As $\{\xi_i\}$ is a monotonically decreasing sequence, it follows that 
$A_m(j)$ is a monotonically decreasing function of $j$.
With these definitions in hand, we are now ready to construct our adversary
matrix.

\begin{proof}[Proof of \thmref{thm:ospadv} (lower bound on non-negative adversary)]
We first consider the case where $n=2m$ is even.  In \eqnref{eq:primal}, let
$$
  \g_i = A_m(i-1)-A_m(i).
$$
As $A_m(j)$ is a monotonically decreasing function of $j$, we have
$\g_i \ge 0$.  Also note that $A_m(i)=0$ for $i \ge m$, so 
$\Toeplitz(\g_n,\ldots, \g_1)$ is bipartite.

The objective function is a telescoping series, so the value of the semidefinite program is
$$
  2 A_m(0) = 2 \sum_{i=0}^{m-1} \xi_i^2,
$$
as claimed.  Thus it suffices to show that $\norm{\Toeplitz(\g_n,\ldots,\g_1)} \le 1$.

We will show that, in fact, $\norm{\Toeplitz(\g_n,\ldots,\g_1)} = 1$.  We do this by exhibiting an
eigenvector $u$ with eigenvalue $1$, and with strictly positive entries.  This will finish the proof by the following argument:
As $\Toeplitz(\g_n,\ldots, \g_1)$ is a non-negative, symmetric matrix, its spectral norm is equal to its largest eigenvalue.  By the Perron-Frobenius theorem, it has a principal eigenvector with non-negative entries.  As the eigenvectors of a symmetric matrix corresponding to distinct eigenvalues are orthogonal, and no non-negative vector can be orthogonal to $u$, we conclude that the largest eigenvalue must agree with the eigenvalue of $u$, and so is $1$.

The relevant eigenvector of $\Toeplitz(\g_n,\ldots,\g_1)$ is
\be
 u := (\xi_0, \xi_1, \ldots, \xi_{m-1}, \xi_{m-1}, \ldots, \xi_1, \xi_0).
\label{eq:ueven}
\ee
Computing $\Toeplitz(g_n,\ldots,g_1)u$, we see that $u$ is an eigenvector with eigenvalue $1$ provided
\be
  \sum_{i=0}^{m-j-1} \big(A_m(i+j)-A_m(i+j+1)\big) \xi_{i} = \xi_j
\label{eq:evcond}
\ee
for each $j=0,1,\ldots,m-1$.

We give two proofs of \eqnref{eq:evcond}.  In the first proof, we use generating functions.  Define a complementary function to $g(z)$, namely the polynomial
$$
  h(z) := \xi_{m-1} + \xi_{m-2} z + \ldots + \xi_0 z^{m-1},
$$
and consider the product $g(z) h(z)$.  For $i=0,\ldots,m-1$, the coefficient of $z^i$ in this series is $A_m(m-i-1)$, so the coefficent of $z^i$ in $(1-z) g(z) h(z)$ is $A_m(m-i-1)-A_m(m-i)$.  Thus the coefficient of $z^{m-j-1}$ in $(1-z) g(z) h(z) g(z)=h(z)$ is the left hand side of \eqnref{eq:evcond}.  But the coefficient of $z^{m-j-1}$ in $h(z)$ is the coefficient of $z^j$ in $g(z)$, which is simply $\xi_j$, the right hand side of \eqnref{eq:evcond}.

Alternatively, we can explicitly expand the left hand side of \eqnref{eq:evcond}, giving
\bas
\sum_{i=0}^{m-j-1} \big(A_m(i+j)-A_m(i+j+1)\big) \xi_{i} &=
\sum_{i=0}^{m-j-1} \Bigg(\sum_{k=0}^{m-(i+j)-1}\xi_{k}\xi_{k+i+j} \xi_i -
\sum_{k=0}^{m-(i+j)-2}\xi_k \xi_{k+i+j+1}\xi_i \Bigg) \\
&=\sum_{s=0}^{2m-j-1} \sum_{i=0}^{s} \xi_{s-i}\xi_i \xi_{s+j} -
\sum_{s=0}^{2m-j-2} \sum_{i=0}^s \xi_{s-i}\xi_i \xi_{s+j+1} \\
&= \xi_j,
\eas
where in the last step we have used \propref{prop:binom}.  This completes the proof when $n$ is even.

For $n=2m+1$ odd, let
$$
  \g_i = \frac{1}{2} \big(A_{m+1}(i-1) - A_{m+1}(i) 
                        + A_m(i-1)     - A_m(i)\big).
$$
Then the objective function is
$$
  A_{m+1}(0) + A_m(0)
  = 2 \sum_{i=0}^{m-1} \xi_i^2 + \xi_m^2
$$
as claimed.  Now it suffices to show that
\be
  u := (\xi_0, \xi_1, \ldots, \xi_{m-1}, \xi_m, \xi_{m-1}, \ldots, \xi_1, \xi_0)
\label{eq:uodd}
\ee
is an eigenvector of $\Toeplitz(\g_n,\ldots,\g_1)$ with eigenvalue $1$.
(Note that for $n$ odd, $\Toeplitz(\g_n,\ldots,\g_1)$ is irreducible, so $u$ is actually the unique principal eigenvector.)
For all but the middle component of the vector $\Toeplitz(\g_n,\ldots,\g_1) u$, the required condition is simply the average of \eqnref{eq:evcond} and the same equation with $m$ replaced by $m+1$.  For the middle component, we require $A_{m+1}(m) \xi_0 = \xi_m$, which holds because $A_{m+1}(m)= \xi_0 \xi_m$ and $\xi_0=1$.
\end{proof}

\begin{figure}
\centering
\includegraphics[scale=.75]{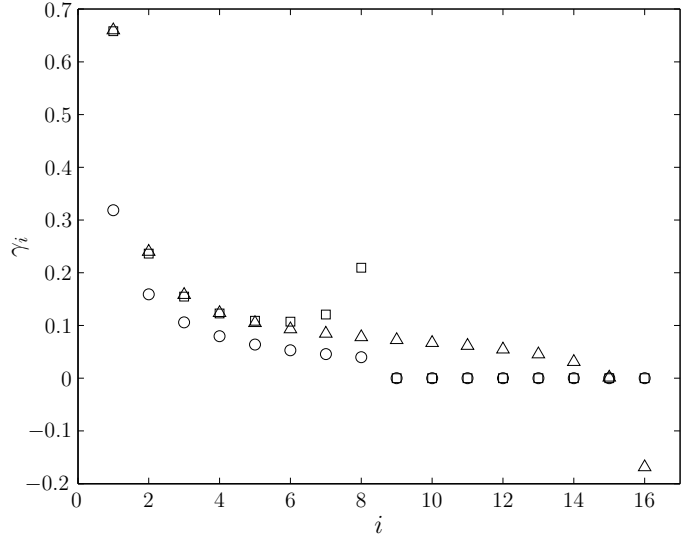}
\caption{Comparison of the weights $\gamma_i$ with $n=16$ for various adversary bounds: the bound of H{\o}yer, Neerbek, and Shi (circles), the optimal non-negative adversary (squares), and the optimal negative adversary (triangles).}
\label{fig:weights}
\end{figure}

In the bound of H{\o}yer, Neerbeck, and Shi, the weight given to a pair $(x,y)$ is inversely proportional to the Hamming distance between $x$ and $y$.  This follows the intuition that pairs which are easier for an adversary to distinguish should be given less weight.  It is interesting to note that the optimal weight scheme does \emph{not} have this property---indeed, at large Hamming distances the weights actually increase with increasing Hamming distance, as shown in \figref{fig:weights}.

\subsection{Dual}

We now show that this bound is optimal by giving a matching solution to the dual semidefinite program \eqref{eq:dual}.

\begin{proof}[Proof of \thmref{thm:ospadv} (upper bound on non-negative adversary)]
Fix $n$, and let $u$ be the vector of length $n$ defined by \eqnref{eq:ueven} if $n$ is even, or by \eqnref{eq:uodd} if $n$ is odd.  Notice that in either case, $u_i=u_{n-i+1}$.  
Let $P = uu^T$, a rank one matrix.  This matrix is positive semidefinite, and its trace is $\norm{u}^2$,
which matches the value of our solution to the primal problem in the previous section.
Thus it suffices to verify that $\Tr_i(P) \ge 1$.  We have
\bas
\Tr_i(P) &=\sum_{j=1}^{n-i} P[j,i+j]
          =\sum_{j=1}^{n-i} u_j u_{i+j} 
          =\sum_{j=1}^{n-i} u_j u_{n-i-j+1}.
\eas
Since $\{\xi_i\}$ is monotonically decreasing in $i$, we have $u_j \ge \xi_{j-1}$, with equality holding when $j \le \ceil{n/2}$.  Thus
$$
  \Tr_i(P) \ge \sum_{j=1}^{n-i} \xi_{j-1} \xi_{n-i-j}= 1
$$
by \propref{prop:binom}.  When $i>\floor{n/2}$, this inequality holds with equality.
\end{proof}

Having established the optimal adversary bound for $\OSP_n$, let us examine its asymptotic behavior.
\begin{corollary}
\label{cor:adv_est}
$$
  \ADV(\OSP_n) = \frac{2}{\pi} (\ln n + \gamma + \ln 8) + O(1/n)
$$
where $\gamma \approx 0.577$ is the Euler-Mascheroni constant.
\end{corollary}
\noindent

\begin{proof}
The generating function for the sequence $\{\xi_i^2\}$ is ${}_2F_1(\frac{1}{2},\frac{1}{2};1;z) = \frac{2}{\pi} K(z)$, where $K(z)$ is the complete elliptic integral of the first kind.  Thus the generating function for $\{\ADV(\OSP_{2m})\}$, which is twice the $m$th partial sum of $\{\xi_i^2\}$, is $\frac{4}{\pi}K(z)/(1-z)$.  The function $K(z)$ is analytic for $|z|<1$, and can be analytically continued to the rest of the complex plane, with the only singularities consisting of branch points at $z=1$ and $z=\infty$ \cite{Olv74}*{Sec.~5.9.1}.  In particular, the logarithmic singularity at $z=1$ has the expansion \cite{BF54}*{Eq.~900.05}
$$
  K(z)
  = \ln \frac{4}{\sqrt{1-z}}
  + \frac{1}{4} (1-z) \left(\ln \frac{4}{\sqrt{1-z}} - 1\right)
  + O\left((1-z)^2 \ln\frac{1}{1-z}\right).
$$
Now let $[z^m] f(z)$ denote the coefficient of $z^m$ in $f(z)$.  According to Darboux's method (see for example \cite{Olv74}*{Sec.~8.9}), we have
\bas
  \ADV(\OSP_{2m})
  &= [z^m] \frac{4}{\pi} \cdot \frac{K(z)}{1-z} \\
  &= [z^m] \frac{4}{\pi} \left(
       \frac{1}{1-z} \cdot \frac{1}{2} \ln \frac{1}{1-z}
     + \frac{\ln 4}{1-z}
     \right) + O(1/m) \\
  &= \frac{2}{\pi}(\ln m + \gamma + \ln 16) + O(1/m),
\eas
where we have used \cite{Jun31}
\bas
  [z^m] \frac{1}{1-z} \ln\frac{1}{1-z} &= \ln m + \gamma + O(1/m)
\eas
and the facts that $[z^m](1-z)^{-1}=1$ and $[z^m]\ln\frac{1}{1-z} = 1/m$.  This proves the corollary for $n=2m$ even.  For $n=2m+1$ odd, we have $\ADV(\OSP_{2m+1})=\ADV(\OSP_{2m}) + \xi_m^2$, and it suffices to observe that $\xi_m^2 = O(1/m)$ by Stirling's approximation.
\end{proof}

\begin{figure}
\centering
\includegraphics[scale=.75]{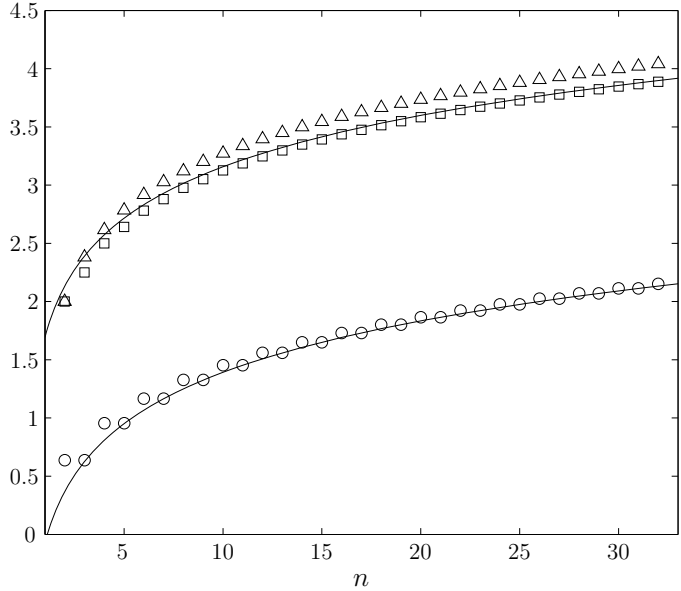}
\caption{Comparison of adversary lower bounds for ordered search: the bound of H{\o}yer, Neerbek, and Shi (circles), the optimal non-negative adversary (squares), and the optimal negative adversary (triangles).  The lower curve shows the asymptotic approximation $\frac{2}{\pi}(\ln n + \gamma - \ln 2)$ of the H{\o}yer-Neerbek-Shi bound, and the upper curve shows the asymptotic approximation $\frac{2}{\pi}(\ln n + \gamma + \ln 8)$ of the non-negative adversary.}
\label{fig:bounds}
\end{figure}

For comparison, the bound of H{\o}yer, Neerbek, and Shi for $\OSP_n$ is $\frac{2}{\pi} H_{\floor{n/2}} = \frac{2}{\pi} (\ln n + \gamma - \ln 2) + O(1/n)$, where $H_n := \sum_{i=1}^n \frac{1}{i}$ is the $n$th harmonic number.
(Note that for the original, unsymmetrized ordered search problem treated in \cite{HNS02}, the bound is $\frac{2}{\pi}(H_n - 1)$.)
Indeed, the optimal value of the non-negative adversary is considerably better for small values of $n$, as shown in \figref{fig:bounds}.

\section{Negative adversary}
\label{sec:neg}

We now turn to the negative adversary method, and give an upper bound on $\MADV(\OSP_n)$ by exhibiting a solution to \eqref{eq:neg_dual}.

Notice that if we find a symmetric matrix $R$ such that $\Tr_i(R)=1$ for $i=0,\ldots,n-1$, we can translate this into a solution to $\eqref{eq:neg_dual}$ by decomposing $R=P-Q$ as the difference of two positive semidefinite matrices with disjoint support, letting $P$ be the projection of $R$ onto its positive eigenspace and letting $Q$ be the projection of $R$ onto its negative eigenspace.  In this case, $\Tr(P+Q)$ is simply $\trn{R}$, the sum of the absolute values of the eigenvalues of $R$.

In looking for a matrix $R$ satisfying $\Tr_i(R)=1$ for all $i$, a natural starting point is our solution to the non-negative dual \eqref{eq:dual}.  Recall that in this construction, for $i>\ceil{n/2}$, the condition $\Tr_i(P)=1$ held with equality.  We imitate that construction by letting
$$
  R[i,j] =
  \begin{cases}
    \xi_i \xi_{n-j+1} & i \le j \\
    \xi_{n-i+1} \xi_j & i > j.
  \end{cases}
$$
Above the diagonal, $R$ looks like a rank one matrix, but it is symmetrized below the diagonal.  
By the convolution property of the $\xi_i$'s we see that $\Tr_i(R)=1$ for $i=0,\ldots,n-1$.  

To upper bound the trace norm of $R$, the following lemma will be helpful:

\begin{lemma}
\label{lem:symrankone}
Let $M$ be an $n \times n$ matrix with entries
\be
  M[i,j] =
  \begin{cases}
    v_i w_j & i \le j \\
    v_j w_i & i > j
  \end{cases}
\label{eq:symrankone}
\ee
where the vectors $v,w \in \R^n$ have positive components, and satisfy $\frac{v_i}{v_{i+1}} > \frac{w_i}{w_{i+1}}$ for $i=1,\ldots,n-1$.  Then $M$ has one positive eigenvalue and $n-1$ negative eigenvalues, and its trace norm satisfies
$$
  2 \norm{v}\norm{w}-v \cdot w
  \le
  \trn{M}
  \le
  2\norm{v}\norm{w}+v \cdot w.
$$
\end{lemma}

\begin{proof}
Sylvester's law of inertia states that the triple of the number of positive, zero, and negative eigenvalues of a matrix
$M$ and that of a matrix $SMS^T$ are the same, provided $S$ is non-singular.  
We apply this law with $S$ given by the $n \times n$ upper tridiagonal matrix with entries
$$
  S[i,j] =
  \begin{cases}
    1        & i=j \\
    -v_i/v_j & i=j-1 \\
    0        & \text{otherwise}.
  \end{cases}
$$
Then a straightforward calculation shows that $SMS^T$ is diagonal, with entries
$$
  (SMS^T)[i,i] =
  \begin{cases}
    \frac{v_i}{v_{i+1}}(v_{i+1} w_i - v_i w_{i+1}) & i=1,\ldots,n-1 \\
    v_n w_n                                        & i=n.
  \end{cases}
$$
By the assumptions of the lemma, the first $n-1$ diagonal entries are negative, and the last is positive; thus $M$ has one positive eigenvalue and $n-1$ negative eigenvalues.

As $M$ is a symmetric, non-negative matrix, its positive eigenvalue is equal to $\sn{M}$, so $\trn{M}+\Tr(M) = 2 \sn{M}$.   Notice that $\Tr(M) = v \cdot w$.  Because $M$ is non-negative and entrywise larger than the rank one matrix $vw^T$, we have $\sn{M} \ge \norm{vw^T} = \norm{v}\norm{w}$.  Furthermore, because $M$ is entrywise smaller than the rank two matrix $A=vw^T + wv^T$, we have $\sn{M} \le \norm{vw^T + wv^T}$.  Using the facts that $\Tr(A) = \lambda_1(A)+\lambda_2(A)=2v\cdot w$ and that $\Tr(A^2) = \lambda_1(A)^2 + \lambda_2(A)^2 = 2(\norm{v}^2 \norm{w}^2 + (v\cdot w)^2)$, we see that the eigenvalues of $A$ are $v \cdot w \pm \norm{v}\norm{w}$.  Thus we conclude that $\norm{M} \le \norm{v}\norm{w} + v \cdot w$, and the lemma follows.
\end{proof}

Now we are ready to finish the proof of \thmref{thm:ospadv}. 

\begin{proof}[\thmref{thm:ospadv} (upper bound on negative adversary)]
The matrix $R$ defined above is of the form \eqnref{eq:symrankone} with $v=(\xi_0,\xi_1,\ldots,\xi_{n-1})$ and $w=(\xi_{n-1},\xi_{n-2},\ldots,\xi_0)$, the reversal of $v$.  By \propref{prop:binom}, $\Tr_i(R) = 1$ for $i=0,\ldots,n-1$, so $R$ is a solution of \eqnref{eq:neg_dual}.  Since $v$ is monotonically increasing and $w$ is monotonically decreasing, the conditions of \lemref{lem:symrankone} are satisfied, and thus $\trn{R}\le 2\norm{v}^2+1=\ADV(\OSP_{2n})+1$.

Finally, using \corref{cor:adv_est} we find
$$
  \ADV(\OSP_{2n}) - \ADV(\OSP_n) \le \frac{2}{\pi}\ln 2 + O(1/n),
$$
so 
\bas
  \MADV(\OSP_n) &\le \ADV(\OSP_n) + 1 + \frac{2}{\pi}\ln 2 + O(1/n). \qedhere
\eas
\end{proof}

Note that the solution of \eqnref{eq:neg_dual} given above is not the optimal one.
For fixed $n$, we can find the optimal solution using a numerical semidefinite program solver.  \figref{fig:weights} shows the optimal weights for $n=16$, and \figref{fig:bounds} shows the value of the optimal negative adversary bound for $n=2$ through $32$.
Empirically, we have found that in the optimal solution, $P-Q$ is a rank two matrix in which $P,Q$ are of the form
$$
  P = p p^T \,,~ Q = q q^T
  \quad\text{with}\quad
  p_i = r_i \cos\theta_i \,,~ q_i = r_i \sin\theta_i,
$$
where
$$
  \theta_i = \frac{\pi}{2n-1}\left(\frac{n+1}{2}-i\right)
$$
and
$$
  r_i^2 \approx \begin{cases}
    \frac{1}{n+1} \csc\frac{1}{(n+1) \xi_{i-1}^2} &
      i=1,\ldots,\ceil{n/2} \\
    r_{n-i+1}^2 & i=\ceil{n/2}+1,\ldots,n.
  \end{cases}
$$
However, we do not know the exact form of $r$ or the optimal negative adversary value.  

\section{Conclusion}
\label{sec:conclusion}

We have given upper bounds on the lower bounds provable by the quantum 
adversary method for the ordered search problem, showing that both the
standard and negative adversary values are $\frac{2}{\pi} \ln n + O(1)$.  In particular, we have shown that establishing the quantum query complexity of ordered search will either require a lower bound proved by a different technique, or an improved upper bound.  On the lower bound side, one could investigate the bounds given by the polynomial method \cite{BBCMW01}, or by the recently developed multiplicative adversary technique of {\v S}palek \cite{Spa07}.  However, we feel that it is more likely that the $\frac{1}{\pi} \ln n$ lower bound is in fact tight, and that further improvement will come from algorithms.  As the current best upper bounds are ad hoc, based on numerical searches, they can almost certainly be improved.

The disagreeable reader may argue that upper bounds on lower bounds are only meta-interesting.  We counter this objection as follows.  Barnum, Saks, and Szegedy have exactly characterized quantum query complexity in terms of a semidefinite program \cite{BSS03}.  The adversary method can be viewed as a  relaxation of this program, removing some constraints and  focusing only on the output condition.  Thus, our results can be viewed as solving a simplification of the quantum query complexity semidefinite program, which might provide insight into the solution of the full program.  Indeed, we hope that the results presented here will be a useful step toward determining the quantum query complexity of ordered search.

\section*{Acknowledgments}

We thank Peter H{\o}yer for stimulating discussions, and in particular, for suggesting the second proof of Equation~\eqref{eq:evcond} presented above.
This work was done in part while AMC was at the Institute for Quantum 
Information at the California Institute of Technology, where he received 
support from NSF Grant PHY-0456720 and ARO Grant W911NF-05-1-0294.  TL is 
supported by a Rubicon grant from the Netherlands Organisation for Scientific 
Research (NWO)  and by the European Commission under the Integrated Project 
Qubit Applications (QAP) funded by the IST directorate as Contract Number 
015848.


\end{document}